\documentclass[aps,twocolumn,showpacs]{revtex4}
\usepackage{graphicx}
\usepackage{amsmath}
\usepackage{amsfonts}
\usepackage{color}
\newcommand{\tbox}[1]{\mbox{\tiny #1}}

\begin{document}



\title{Scaling properties of the action in the Riemann-Liouville 
fractional standard map}


\author{J. A. M\'endez-Berm\'udez}
\address{Instituto de F\'isica, Benem\'erita Universidad Aut\'onoma de Puebla, 
Puebla 72570, Mexico}

\author{R. Aguilar-S\'anchez}
\address{Facultad de Ciencias Qu\'imicas, Benem\'erita Universidad Aut\'onoma de Puebla,
Puebla 72570, Mexico}

\author{Jos\'e M. Sigarreta}
\address{Facultad de Matem\'aticas, Universidad Aut\'onoma de Guerrero, 
Carlos E. Adame No.54 Col. Garita, Acalpulco Gro. 39650, Mexico}

\author{Edson D. Leonel}
\address{Universidade Estadual Paulista (UNESP) - Departamento de F\'isica,
Av. 24A, 1515 -- Bela Vista -- CEP: 13506-900 -- Rio Claro -- SP -- Brazil}

\begin{abstract}
The Riemann-Liouville fractional standard map (RL-fSM) is a two-dimensional nonlinear map 
with memory given in action-angle variables $(I,\theta)$.  
The RL-fSM is parameterized by $K$ and $\alpha\in(1,2]$ which control the strength of nonlinearity 
and the fractional order of the Riemann-Liouville derivative, respectively.
In this work, we present a scaling study of the average squared action $\left< I^2 \right>$ of the RL-fSM
along strongly chaotic orbits, i.e.~for $K\gg1$.
We observe two scenarios depending on the initial action $I_0$, $I_0\ll K$ or $I_0\gg K$.
However, we can show that $\left< I^2 \right>/I_0^2$ is a universal function of the 
scaled discrete time $nK^2/I_0^2$ ($n$ being the $n$th iteration of the RL-fSM).
In addition, we note that $\left< I^2 \right>$ is independent of $\alpha$ for $K\gg1$.
Analytical estimations support our numerical results.
\end{abstract}

\maketitle

\section{Preliminaries}

The kicked rotor represents a free rotating stick in an inhomogeneous field that is periodically switched on 
in instantaneous pulses, see e.g.~\cite{O08}. It is described by the second order differential equation
\begin{equation}
    \ddot{\theta} + K \sin(\theta) \sum_{j = 0}^\infty \delta\left(\frac{t}{T} - j \right) = 0 .
    \label{KR}
\end{equation}
Here, $\theta\in [0,2\pi]$ is the angular position of the stick, $K$ is the kicking strength, $T$ is the kicking 
period (that we set to one from now on), and $\delta$ is Dirac's delta function.
By replacing the second-order derivative in the equation of motion of the kicked rotor with a 
Riemann-Liouville (RL) derivative of fractional order $\alpha$~\cite{SKM93,KST06}, the RL fractional kicked 
rotor (fKR) is obtained~\cite{TZ08,ET09}:
\begin{equation}
_0D_t^\alpha \theta + K \sin(\theta) \sum_{j = 0}^\infty \delta\left(t - j \right) = 0 , \quad 1< \alpha \leq 2 .
\label{fKR}
\end{equation}
Above~\cite{SKM93,KST06},
\begin{align}
&_0D_t^\alpha \theta(t) =  D_t^m {_0 {\cal I}}_t^{m-\alpha}\theta(t) \nonumber \\ 
& = \frac{1}{\Gamma(m-\alpha)}\frac{d^{m}}{dt^{m}}\int_{0}^{t}\frac{\theta^{\tau}d\tau}{(t-\tau)^{\alpha-m+1}}, \quad m-1<\alpha\leq m, \nonumber
\end{align}
with $D_t^m = d^m/dt^m$, $_0{\cal I}_t^m f(t)$ is a fractional integral given by
\begin{equation*}
_0{\cal I}_t^m f(t) = \frac{1}{\Gamma(m)}\int_0^t (t-\tau)^{\alpha-1}f(\tau)d\tau ,
\end{equation*}
and $\Gamma$ is the Gamma function.

The RL-fKR has a stroboscopic version, a two-dimensional nonlinear map with memory, which is 
well known as the RL fractional standard map (RL-fSM)~\cite{ET09}:
\begin{equation}
\begin{array}{ll}
I_{n+1} = I_n - K\sin(\theta_n) , \\
\theta_{n+1} = \displaystyle{ \frac{1}{\Gamma(\alpha)}\sum_{i = 0}^{n} I_{i+1} V^1_\alpha(n-i+1)}, \quad \mbox{mod}~(2\pi),
\end{array}
\label{RLfSM}
\end{equation}
where $I(t) \equiv {_0D}_t^{\alpha-1}\theta(t)$, $n$ is the discrete time, and
$V^k_\alpha(m) = m^{\alpha-k}-(m-1)^{\alpha-k}$.
Then, the RL-fSM, given in action-angle variables $(I,\theta)$, is parameterized by $K$ and $\alpha\in(1,2]$ 
which control the strength of nonlinearity and the fractional order of the RL derivative, respectively.
In fact, for $\alpha=2$, the RL-fSM reproduces the celebrated Chirikov's standard map (CSM)~\cite{C69}.

Compared with CSM, which presents the generic transition to chaos (in the context of 
Kolmogorov--Arnold--Moser theorem), depending on the parameter pair ($K,\alpha$),
the RL-fSM shows richer dynamics: It generates attractors (fixed points, asymptotically 
stable periodic trajectories, slow converging and slow diverging trajectories, ballistic 
trajectories, and fractal-like structures) and/or chaotic trajectories~\cite{ET09,ET13,E19,E11,MPSL23}. 
Moreover, trajectories may intersect and attractors may overlap~\cite{E11}.

Among the available studies on the RL-fKR, see e.g.~\cite{ET09,ET13,E19,E11,MPSL23}, the 
analysis of strongly chaotic orbits has been left unexplored.
Therefore, here we undertake this task and characterize the dynamics of the RL-fSM by computing 
the squared average action $\left< I^2_n \right>$ 
when $K\gg1$.

\begin{figure*}[ht]
\centering
\includegraphics[width=0.9\textwidth]{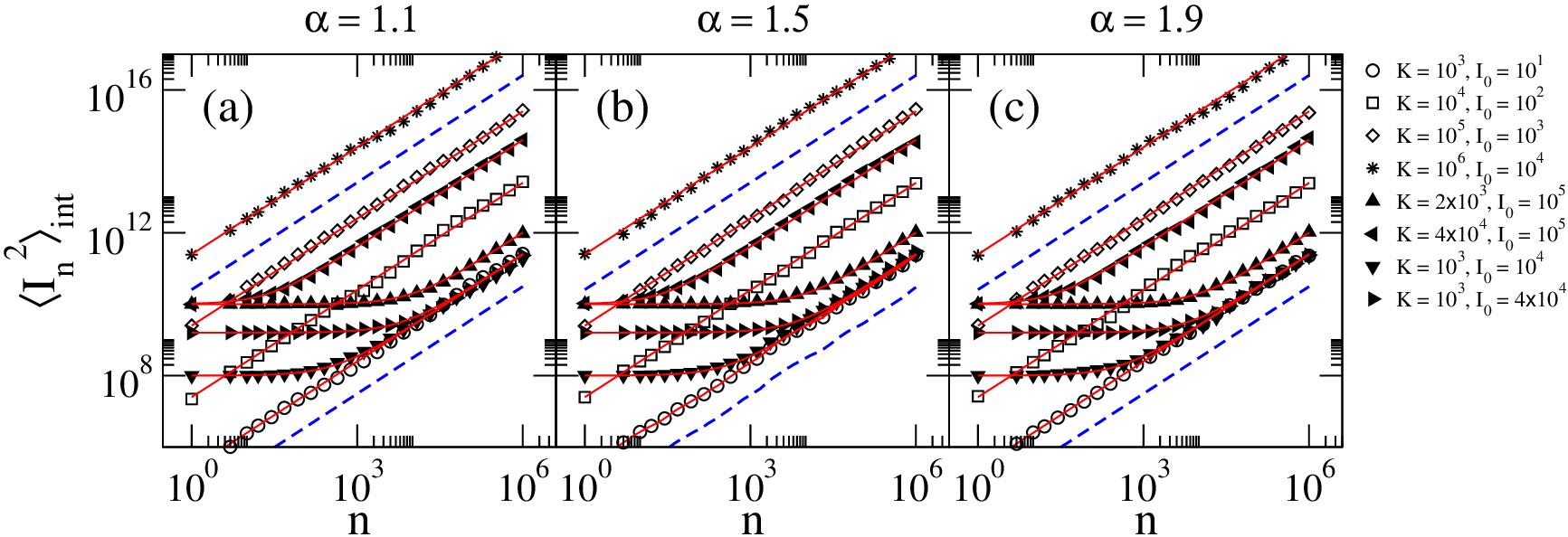}
\caption{Average squared action $\left< I^2_n \right>_{\tbox{int}}$ as a function of the discrete time $n$ for 
(a) $\alpha=1.1$, (b) $\alpha=1.5$, and (c) $\alpha=1.9$. 
Open symbols (full symbols) correspond to $I_0 \ll K$ ($I_0 \gg K$). 
The blue dashed lines, plotted to guide the eye, are proportional to $n$.
The average is over 100 orbits with initial random phases in the interval $0<\theta_0<2\pi$. 
Red full lines are Eq.~(\ref{I2g0RL}).}
\label{Fig01}
\end{figure*}
\begin{figure*}[ht]
\centering
\includegraphics[width=0.6\textwidth]{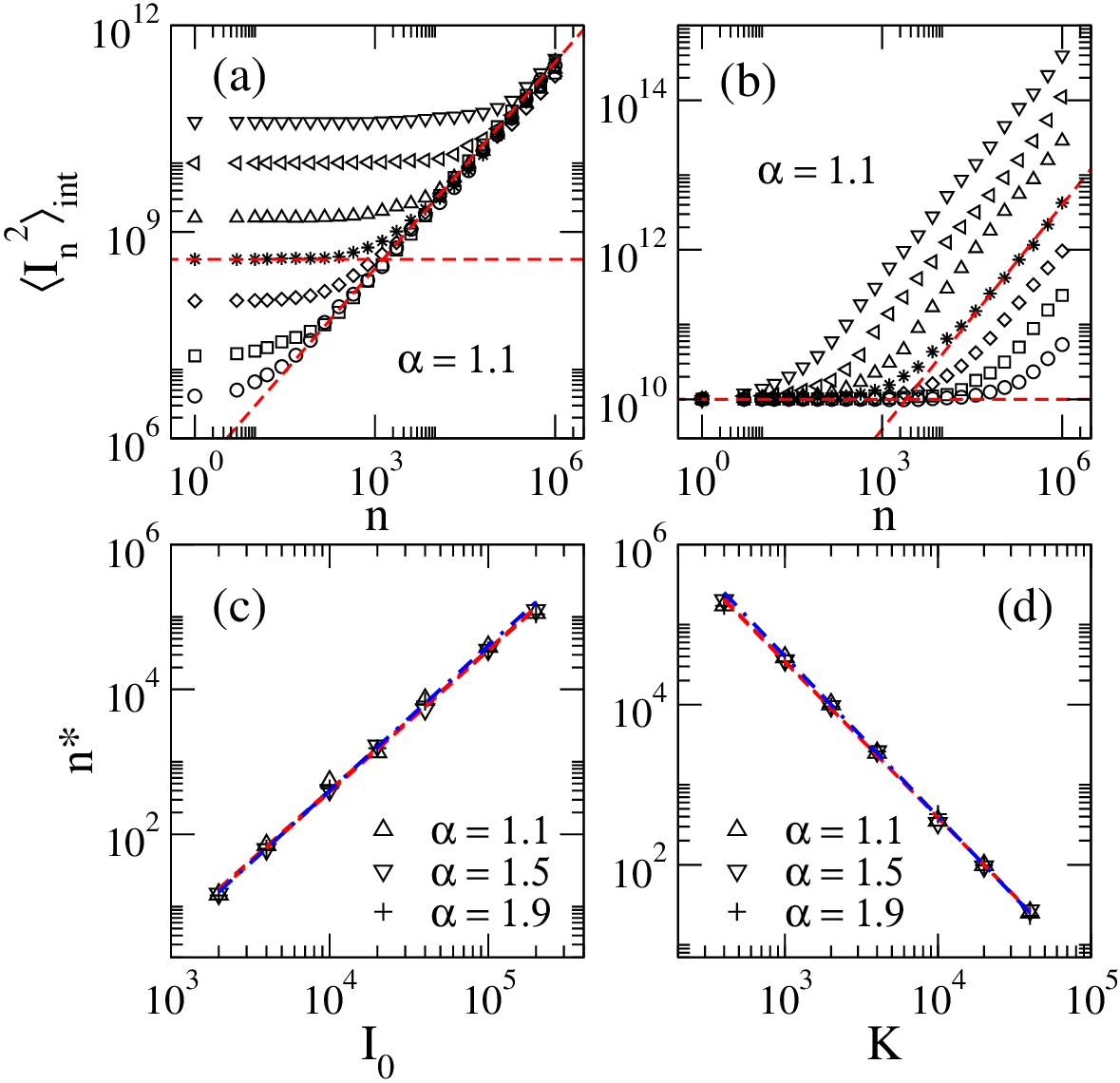}
\caption{(a,b) Average squared action $\left< I^2_n \right>_{\tbox{int}}$ as a function of $n$
for (a) $K=10^3$ and several values of $I_0$ ($2\times 10^3$, $4\times 10^3$, $10^4$, $2\times 10^4$,
$4\times 10^4$, $10^5$, and $2\times 10^5$, from bottom to top) and (b) $I_0=10^5$ and several values of 
$K$ ($4\times 10^2$, $10^3$, $2\times 10^3$, $4\times 10^3$, $10^4$, $2\times 10^4$, and $4\times 10^4$,
from bottom to top). 
As in Fig.~\ref{Fig01}, the average is taken over 100 orbits with initial random phases in the interval 
$0<\theta_0<2\pi$. 
All data in (a,b) correspond to $\alpha = 1.1$. 
The horizontal red dashed lines in (a,b) indicate $\left< I^2_n \right>_{\tbox{int}}=I_0^2$ with (a) 
$I_0=4\times 10^4$ and (b) $I_0=10^5$, respectively.
The transversal red dashed lines in (a,b) are fittings of $\left< I^2_n \right>_{\tbox{int}}= {\cal C} n$ 
to the data represented by asterisks (for $n\ge10^4$) with fitting constants (a) ${\cal C}=294658$ 
and (b) ${\cal C}=3984348$.
(c,d) Crossover time $n^*$  (a) as a function of $I_0$ for constant $K$ ($K=10^3$) and (b) as a function of 
$K$ for constant $I_0$ ($I_0=10^5$). In (c,d), three values of $\alpha$ are reported: $\alpha=1.1$, 1.5, and 1.9.
Red dashed lines in (c,d) are power-law fittings to the data of the form (a) $n^*\propto I_0^{\gamma_1}$ with 
$\gamma_1\approx 2$ and (b) $n^*\propto K^{\gamma_2}$ with $\gamma_2\approx -2$.
Blue dot-dashed lines in (c,d) are Eq.~(\ref{nCO}).}
\label{Fig02}
\end{figure*}
\begin{figure*}[ht]
\centering
\includegraphics[width=0.9\textwidth]{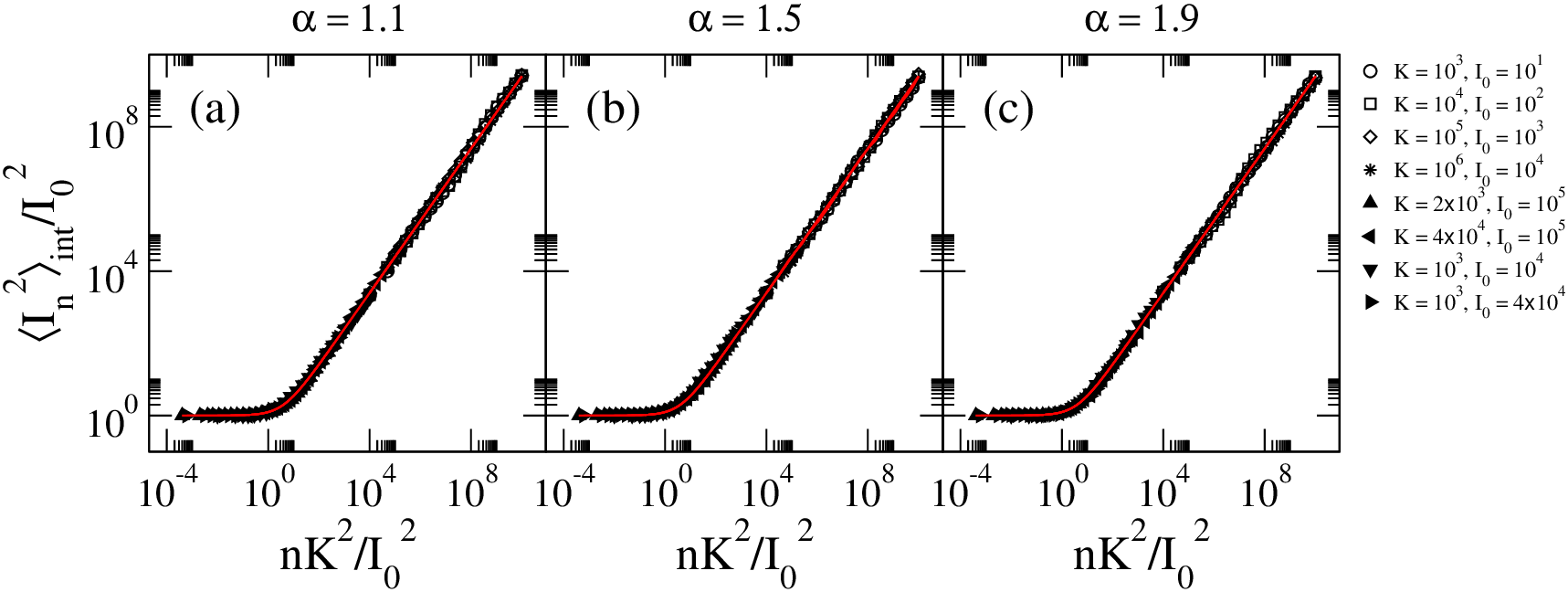}
\caption{Normalized average squared action $\left< I^2_n \right>_{\tbox{int}}/I_0^2$ as a function 
of the normalized time $nK^2/I_0^2$. Same data sets as in Fig.~\ref{Fig01}.
The red line is Eq.~(\ref{I2g0scaling}).}
\label{Fig03}
\end{figure*}

\section{Numerical results}
\label{numericsI2}

For the numerical study we wrote a program in FORTRAN90 to compute the orbits of  
map~(\ref{RLfSM}) by straightforward iteration. Moreover, since the memory property in 
map~(\ref{RLfSM}) forbids the use 
of a large number of orbit realizations, instead of
investigating $\left< I^2_n \right>$ directly, to smooth the curves $\left< I^2_n \right>$ vs.~$n$, 
we compute its cumulative-normalized value
\begin{equation}
\left< I^2_n \right>_{\tbox{int}} = \frac{1}{n} \int^n_{n_0=0} \left< I^2_{n'} 
\right> dn'.
\label{I2int}
\end{equation}
Specifically, we compute $\left< I^2_n \right>_{\tbox{int}}$ for map~(\ref{RLfSM}) following two steps: 
First we calculate the average squared action over the orbit associated with the
initial condition $j$ as
$\left< I^2_{n,j} \right> = (n+1)^{-1} \sum^n_{i=0} I^2_{i,j}$, 
where $i$ refers to the $i$-th iteration of the map. Then, $\left< I^2_n \right>_{\tbox{int}}$ is defined 
as the average over $M=100$ independent realizations of the map (by randomly choosing values of 
$\theta_0$ in the interval $0<\theta_0<2\pi$):
$\left< I^2_n \right>_{\tbox{int}}(I_0,K,\alpha) = M^{-1} \sum^M_{j=1} \left< I^2_{n,j} \right>$.

In Fig.~\ref{Fig01} we plot $\left< I^2_n \right>_{\tbox{int}}$ as a function of the discrete time $n$ 
for representative values of $\alpha$ in the interval $(1,2]$: (a) $\alpha=1.1$, (b) $\alpha=1.5$, and (c) $\alpha=1.9$. Several 
combinations of parameter pairs $(I_0, K)$ are considered, as indicated in the r.h.s. of the figure.
From this figure, we observe two scenarios depending on the initial action $I_0$ as compared with
$K$: $I_0\ll K$ (open symbols) or $I_0\gg K$ (full symbols).
Specifically, when $I_0\ll K$, $\left< I^2_n \right>_{\tbox{int}}\propto n$ for all $n$ (see the
blue dashed lines). While for $I_0\gg K$, first, $\left< I^2_n \right>_{\tbox{int}}$ remains approximately constant 
and proportional to $I^2_0$ up to a crossover time $n^*$, after which $\left< I^2_n \right>_{\tbox{int}}$ grows
proportional to $n$.

From Fig.~\ref{Fig01}, it can also be seen that the crossover time $n^*$ depends on both $I_0$ and $K$,
while the dependence of $n^*$ with $\alpha$ is not evident. Then, to look for the dependence of $n^*$
on the map parameters, in Fig.~\ref{Fig02}(a) [Fig.~\ref{Fig02}(b)] we plot $\left< I^2_n \right>_{\tbox{int}}$
vs.~$n$ for several values of $I_0$ [$K$] and fixed $K$ [$I_0$]. In both figures, we use $\alpha=1.1$.
We numerically extract $n^*$ as the crossing point between the functions 
$\left< I^2_n \right>_{\tbox{int}}=I_0^2$ and $\left< I^2_n \right>_{\tbox{int}}= {\cal C} n$ (which is the fitting 
to the data in the growing regime); as examples, 
see the horizontal and transversal dashed lines in Figs.~\ref{Fig02}(a,b), respectively.
Thus, in Figs.~\ref{Fig02}(c,d) we plot the obtained values of $n^*$ for $\alpha=1.1$ but also for
$\alpha=1.5$ and 1.9. Figures~\ref{Fig02}(c,d) reveal the power-law dependence
\begin{equation}
n^* \propto I_0^{\gamma_1} K^{\gamma_2}
\label{n*}
\end{equation}
and the independence of $n^*$ on $\alpha$.
Power-law fittings of the data in Figs.~\ref{Fig02}(c,d) provide $\gamma_1\approx 2$ and $\gamma_2\approx -2$,
see red dashed lines.

Equation~(\ref{n*}) together with the observation that $\left< I^2_n \right>_{\tbox{int}}\approx I^2_0$ for
$n<n^*$ allow us to scale the curves $\left< I^2_n \right>_{\tbox{int}}$ vs.~$n$. Indeed, in Fig.~\ref{Fig03}
we plot $\left< I^2_n \right>_{\tbox{int}}/I_0^2$ as a function of the normalized time $nK^2/I_0^2$ 
(i.e.~$n/n^*$) and observe the colapse of all curves on top of a {\it universal} function.

\section{Analytical estimation}
\label{analytics}

Now, to support and better understand the scaling performed above, we derive an 
analytical estimation for $\left< I^2_n \right>_{\tbox{int}}$, see e.g.~\cite{MOL16}.

From the first line of map~(\ref{RLfSM}) we have that
$I^2_{n+1}=I_n^2-2KI_n\sin(\theta_n)+K^2\sin^2(\theta_n)$,
so we can write
\begin{equation*}
\left< I^2_{n+1} \right> = \left< I_n^2 \right> 
- \ 2K \left< I_n \right> \left< \sin(\theta_n) \right> + K^2 \left<\sin^2(\theta_n) \right>.
\end{equation*}
Since for chaotic orbits we can assume that $\left< \sin(\theta_n) \right>=0$, the term
$2K \left< I_n \right> \left< \sin(\theta_n) \right>$
can be eliminated. Therefore,
\begin{equation}
\label{I2bdRLfSM}
\left< I^2_{n+1} \right> = \left< I_n^2 \right>  + \frac{K^2}{2} ,
\end{equation}
where we have used $\left<\sin^2(\theta_n) \right>=1/2$.
Then, by noticing that
\begin{equation*}
\left< I^2_{n+1} \right> - \left< I_n^2 \right> =
\frac{\left< I^2_{n+1} \right> - \left< I_n^2 \right>}{(n+1)-n}
\approx \frac{dJ}{dn} \ ,
\end{equation*}
we rewrite Eq.~(\ref{I2bdRLfSM}) as the first order differential equation:
\begin{equation}
\label{dJdndRLfSM}
\frac{dJ}{dn} = \frac{K^2}{2} ,
\end{equation}
where $J \equiv \left< I_n^2 \right>$. Therefore, by solving~(\ref{dJdndRLfSM}), we can write
\begin{equation}
\label{I2dRLfSM}
\left< I^2_n \right> = I_0^2 + \frac{K^2}{2} n ,
\end{equation}
where we have used $J_0 = \left< I_0^2 \right> = I_0^2$ and $n_0=0$.
Finally, by substituting~(\ref{I2dRLfSM}) into Eq.~(\ref{I2int})
we can also write down an explicit expression for 
$\left< I^2_n \right>_{\tbox{int}}$:
\begin{equation}
\label{I2g0RL}
\left< I^2_n \right>_{\tbox{int}} = I_0^2 + \frac{K^2}{4} n .
\end{equation}

Indeed, Eq.~(\ref{I2g0RL}) reproduces well our numerical data as can be seen in 
Fig.~\ref{Fig01} where we have included Eq.~(\ref{I2g0RL}) as red lines.

\section{Discussion and conclusions}

Given the good correspondence of Eq.~(\ref{I2g0RL}) and the numerical data, it is 
clear that it reproduces the scaling laws reported in Sec.~\ref{numericsI2}, which can
be summarized as
\begin{equation*}
\label{scaling1}
\left< I^2_n \right>_{\tbox{int}} = \left\{
\begin{array}{ll}
\propto K^2 n \ , & \quad \mbox{when} \quad I_0 \ll K  , \\
\left.
\begin{array}{ll}
\approx I_0^2 \ , & n<n^* \\
\propto K^2 n \ , & n>n^*
\end{array} 
\right\}
& \quad \mbox{when} \quad I_0 \gg K  .
\end{array}
\right. 
\end{equation*}
Moreover, Eq.~(\ref{I2g0RL}) can also be used to demonstrate that the ratio 
$\left< I^2_n \right>_{\tbox{int}}/I_0^2$ is a simple {\it universal} function of the variable 
$\overline{n}=n/n^*$:
\begin{equation}
\label{I2g0scaling}
\frac{\left< I^2_n \right>_{\tbox{int}}}{I_0^2} 
= 1 + \overline{n} ,
\end{equation}
where the crossover time $n^*$ is now naturally defined as
\begin{equation}
\label{nCO}
n^* \equiv 4I_0^2K^{-2} ,
\end{equation} 
in agreement with Eq.~(\ref{n*}).
Finally, in Fig.~\ref{Fig02}(c,d) and Fig.~\ref{Fig03} we plot Eq.~(\ref{nCO}) and Eq.~(\ref{I2g0scaling}) 
(see dot-dashed blue lines and red full lines), respectively, and observe an excellent agreement with the 
numerical data. 

It is relevant to notice that for strongly chaotic orbits, $K\gg 1$, the average squared action 
$\left< I^2_n \right>$ for the RL-fSM does not depend on the order $\alpha$ of the fractional derivative.
Indeed, the panorama reported here for $\left< I^2_n \right>$ vs.~$n$ is equivalent to that of 
CSM~\cite{MOL16,LS07} as well as that of the discontinuous standard map~\cite{MOL16,MA12}, 
both with $K\gg 1$.
This could be understood from the analytical estimation of Sec.~\ref{analytics} by noticing that
to obtain the expression for $\left< I^2_n \right>$ we mainly used the first equation of 
map~(\ref{RLfSM}) which does not contain the parameter $\alpha$; i.e.~the property of
memory, parametrized by $\alpha$, is only present in the equation for $\theta$ which is a ciclic
variable. So, when $K\gg 1$, $\left< I^2_n \right>$ must be independent of $\alpha$. 
That is, to observe effects of $\alpha$ on the dynamics of the RL-fSM, $K\sim 1$ should be set, see e.g.~\cite{ET09,ET13,E19,E11,MPSL23}.

We stress that similar studies can be carried out for other types of nonlinearity (not just 
the continuous sine-shaped nonlinearity in the first equation of the RL-fSM) and for other types of 
fractional derivatives. This, in fact, will be the subject of future investigations.

Finally, we want to add that our work falls within the scope of the General Fractional Dynamics 
(GFDynamics), a line of research recently introduced in Ref.~\cite{T21}.

\section*{Acknowledgements}

J.A.M.-B. thanks support from CONAHCyT-Fronteras (Grant No.~425854) 
and VIEP-BUAP (Grant No.~100405811-VIEP2024), Mexico.
The research of J.M.S. is supported by a grant from Agencia Estatal de Investigaci\'on 
(PID2019-106433GB-I00/AEI/10.13039/501100011033), Spain.
E.D.L. acknowledges support from CNPq (No.~301318/2019-0) and FAPESP (No.~2019/14038-6), 
Brazilian agencies.


\end{document}